\documentstyle[multicol,prl,aps]{revtex}

\begin{document}

\title{Relational Reality in Relativistic Quantum Mechanics}
\author{ Rodolfo Gambini$^{1}$ Rafael A. Porto$^{ 1}$}

\address{1. Instituto de F\'{\i}sica, Facultad de
Ciencias, Igu\'a 4225, esq. Mataojo, Montevideo, Uruguay.}
\date{May 20th 2001}

\maketitle

\begin{abstract} 
Up to now it has been impossible to find a realistic
interpretation for the reduction process in relativistic quantum
mechanics. The basic problem is the dependence of the states on the frame
within which collapse takes place. A suitable use of the causal structure
of the devices involved in the measurement process allows us to introduce
a covariant notion for the collapse of quantum states. 
\end{abstract}

\begin{multicols}{2}

Relativistic quantum mechanics does not provide us with a
covariant notion for the collapse of a quantum state in a
measurement process. The basic problem is the dependence of the
states on the frame within which the collapse is stipulated to
occur \cite{Bloch}. On the other hand, it is well known that
measurements of local observables, which commute at space-like
separations, yield the correct covariant probabilities
independently of the Lorentz frame used.  In that sense, it does
not matter if the results of the experiments are described by
different Lorentz observers as different and non-covariant quantum
processes at the level of the states.  The reduction postulate has
been controversial in many ways especially due to its well known
consequences, i.e., the loss of unitarity, the micro-macro world's
division, and the non-local character of the theory that leads,
due to the inclusion of relativity, to non-covariant processes
\cite{Bloch,Aha-A,Ahar-A,Aharo-A}. These problems have led many
physicists to adopt an instrumentalist point of view.  Even if we
assume that the measurement problem has been solved, we still have
to understand its non-covariant character. As was pointed out by
D'Espagnat: "Within the world view we are looking for, a state
should collapse covariantly if it collapses at
all."\cite{Esp,DEsp}. However, as there was not, up to now, a
covariant notion of the reduction process consistent with local as
well as non-local properties he concluded "that {\em even if the
measurement problem is considered as solved}, the conception
according to which {\em the world is made of quantum states} is
not consistent with the whole of our physical knowledge and must
be therefore given up."\cite{Esp,DEsp} Here we shall show that it
is possible to introduce a covariant notion of the reduction
process in accordance to the previous requirement.

Realistic interpretations of the quantum theory have found major
difficulties with the inclusion of relativity.  The main problem
is the lack of a single description of the quantum state. In the
non-relativistic domain, a realistic interpretation already
exists. It was first suggested by Heisenberg \cite{Heis} and
developed by Margenau \cite{Marg} and Jordan \cite{Jord}, and is
known as the real tendency interpretation.  It is important to
remark that it only provides a picture of the reduction process,
but it does not solve the measurement problem. Within this
approach a quantum state is a real entity that characterizes the
disposition of the system, at a given value of the time, to
produce certain events with certain probabilities during the
interaction with a set of macroscopic measurement devices. Due to
the uniqueness of the non-relativistic time, the set of
alternatives among which the system chooses is determined without
ambiguities. In fact, they are simply associated to observables
corresponding to certain decomposition of the identity. For each
value of the time where measurement takes place, the system
coupled with the measurement devices ``makes a decision"
\cite{Jord} and produces events with probabilities given by the
state of disposition of the system. The evolution of this state is
also perfectly well defined. For instance, if we adopt the
Heisenberg picture, evolution is given by a sequence of states of
disposition. The dispositions of the system change during the
measurement processes according to the reduction postulate, and
remain unchanged until the next measurement. Of course, the
complete description is covariant under Galilean transformations.
However, up to now, it has been impossible to extend these
properties to the relativistic domain, and consequently all the
attempts of finding a tentative description of reality based on
standard quantum mechanics have been found incomplete
\cite{Bloch,Aha-A,Ahar-A,Aharo-A}. Hellwig and Krauss (H-K)
proposed a covariant description of the reduction process many
years ago \cite{H-K}, their basic assumption being that the
collapse occurs along the backward light cone of the measurement
event. However, as Aharonov and Albert have shown
\cite{Aha-A,Ahar-A}, their description is not consistent with the
measurement of non-local observables. Even more important is the
fact that it does not allow us to preserve the non-relativistic
interpretation in the evolution of the system as a well-defined
sequence of states of disposition. Indeed, in order to define the
states on a given space-like or light-like surface the H-K
prescription requires the knowledge of all the future measurements
to which the system will be subject.

Hence, so far it has been impossible to have any realistic
interpretation of relativistic quantum mechanics. It is meaningful
to notice that quantum field theory has not been of help for
solving this problem \cite{Aha-A}. In this paper, we are going to
propose a covariant description of the reduction process that will
allow us to preserve a realistic interpretation in the
relativistic domain. However, we shall see that only a relational
kind of realism can be entertained. In order to assign
probabilities to properties of the coupled system, one needs to
identify the set of properties among which the system makes a
decision. In non-relativistic quantum mechanics, the system
coupled with the macroscopical objects chooses among properties
(alternatives) that may always be included among a decomposition
of the identity at a certain time $t_0$. The lack of a unique time
variable in the relativistic domain produces the noticed
ambiguities. Thus, our first objective is the identification of an
intrinsic criterion for the ordering of the alternatives. An
intrinsic order may be introduced by making use of the partial
order of events induced by the causal structure of the theory. Let
us now be more specific. Let us consider an experimental
arrangement of measurement devices, each of them associated with
the measurement of certain property over a space-like region of
space-time at a given proper time. No special assumption is made
about the state of motion of each of them. Indeed different proper
times could emerge from this description due to the different
local reference systems of each device. Thus, we may label each
detector in an arbitrary system of coordinates by an open
three-dimensional region $R_a$, and its four-velocity $u_a$ .One
may introduce a partial order in the following way\cite{Sorkin} :
The instrument $A_{{R_1},{u_1}}$ precedes $A_{{R_2},{u_2}}$ if the
region $R_2$ is contained in the forward light cone of $R_1$.
\footnote{The case where one has devices such that only a portion
of the region $R_2$ is contained in the forward light cone of
$R_1$, leads to a subtler causal structure which has interesting
consequences on the global aspects of the relational tendency
interpretation. The details will be discussed in a forthcoming
paper.} Let us suppose that ${A^0}_{{R},{u}}$ precedes all the
others. In other words, let us assume that all the detectors are
inside the forward light cone coming from this initial condition.
That would be the case, for instance, of the instrument that
prepares the initial wave packet in a two-slit experiment. Then,
it is possible to introduce a strict order without any reference
to a Lorentz time as follows.  Define $S^1$ as the set of
instruments that are preceded only by $A^0$.  Define $S^2$ as the
set of instruments that are preceded only by the set $S^1$ and
$A^0$. In general, define $S^i$ as the set of instruments that are
preceded by the sets $S^j$ with $j <i$ and $A^0$.  Notice that any
couple of elements of $S^i$ is separated by space-like intervals.
This procedure defines a covariant order based on the causal
structure of the devices involved in the measurement process. Now
we have to introduce the operators associated with each device
belonging to the set $S^i$ . The crucial observation is that all
the measurements on $S^i$ can be considered as "simultaneous". In
fact, they are associated with local measurements performed by
each device, and hence represented by a set of commuting
operators.  Since we here intend to stay within the realm of
relativistic quantum mechanics we are going to implement these
operators by noticing that a relativistic system may be considered
as a generally covariant system \cite{Dirac,Hen}. These are
constrained systems invariant under general transformations of the
evolution parameter. The Hamiltonian is a linear combination of
the constraints and the quantum observables are constants of the
motion.  Time is identified with some internal clock variable $T$,
and what is actually measured is not the value of a physical
variable $Q$ for certain value of the parameter ${\tau}$ but the
value $Q(T)$ taken by the physical variable when the clock
variable takes the value $T$.  This procedure, using the clock
variables as the proper time of each device, allows us to describe
all the measurements belonging to the set $S^i$ in terms of a
commuting set of operators (Rovelli's evolving constants) in a
generalized Heisenberg picture (G-H-P) defined on a physical
Hilbert space ${\cal H}_{phys}$ of solutions of the constraint
\cite{Ro,Nos}. The commutativity, and self-adjointness, of the
projectors on a ``simultaneous" set $S^i$, associated to different
local measurement devices, assures that all of them can be
diagonalized on a single option. These conditions insure that the
quantum system has a well defined disposition with respect to the
different alternatives of the set $S^i$. Let us call ${\psi}^0$
the state of the system in the G-H-P prepared by $S^0$. Hence,
after the observation of a set of events\footnote{We are defining
{\it event} as the macroscopical result of the interaction of the
system with the measurement device.} $E= \cup_a{{E}_{{R^1}_a}}$,
each one associated with a local measurement in the region $R_a$
belonging to $S^1$, the state will collapse into ${\psi_E}^1$,
given by the normalized projection of ${\psi}_0$.  The projector
${\cal P}_{E}$, connecting ${\psi}^0$ with ${\psi_E}^1$, is
constructed as a product of local projectors related with each
individual event. This product does not depend on the order due to
the commutation of the projectors. If there is not an event on the
$R_a$ region, in other words if nothing is detected in $R_a$, one
needs to project on the orthogonal complement of each possible
event that may occur in this particular region. It is now clear
the relational character of our approach. The quantum systems keep
a dispositional character with respect to the alternatives
belonging to $S^i$ because they are covariantly defined by an
intrinsic order. The change of these dispositions is also well
defined, because once the interaction with the devices belonging
to $S^i$ has concluded, the state collapses into a projected state
belonging to the Hilbert space ${\cal H}_{phys}$ where the state
of disposition lives. The dispositions are relationally defined
with respect to the set of alternatives given by the measurement
devices. Furthermore, the probabilities and states are computed by
using the standard rules of quantum mechanics based on the
existence of self-adjoint, commutative, local projectors. The
methods developed in \cite{Nos} for the treatment of generally
covariant systems allow the completion of a detailed analysis of
the process sketched above. Covariance follows from the existence
of a natural inner product in the physical space of states ${\cal
H}_{phys}$, such that the local projectors are self-adjoint
operators, and from the unitary implementation of the Lorentz
transformations that insures that the mean value of the projector
is a scalar quantity. Up to now we have not specified the explicit
form of the observables. In fact we have introduced a general
framework applicable for a wide kind of relativistic systems.
\footnote{In a forthcoming paper we shall analyze the relational
description of the measurement process in Quantum Field Theory.}

Let us be more specific and consider the Klein-Gordon quantum
mechanics. In this case we have shown that it is possible to
define a relational position observable that coincides with the
Newton-Wigner operator in the Feshbach-Villars
representation\cite{Nos}. Indeed, it is not difficult to show that
the corresponding local projectors exist and commute, up to a
Compton wavelength. Furthermore, there is a natural covariant
inner product in the physical Hilbert space constructed in
Reference \cite{Nos}. A detailed analysis of the properties of
this observable may be found in the same reference. The physical
Hilbert space is constructed by scalar space-time wave functions
$\psi(z,z^0)$ in the generalized Heisenberg picture, which are
annihilated by the constraint (i.e. solutions of the K-G
equation). These states do not evolve, instead they describe the
whole space-time history of the system. As we have said before,
the time variable should not be identified with $z^0$.  Within
this framework, the evolution is described by relational
observables which in this case are $\hat X(T):= \hat q + {\hat p
T} /{\hat p_0}$ and ${\hat \epsilon}:= \frac{\hat p_0}{\sqrt{{\hat
p}^2+m^2}}$ where ${\hat q},{\hat p},{\hat {p_0}}$ are the
perennials quantum operators associated with the initial position,
momentum and energy of the system, and $T$ the clock variable. The
first observable is the relational position operator which
represents the value of $x$ when $x^0=T$, the second is the sign
of the energy. As we have shown it is possible to construct a
covariant inner product without any reference to any particular
Lorentz time, such that these observables are well defined
self-adjoint operators. Hence it is not difficult to construct
local projectors associate with these observables on each local
reference system in the region ${R^j}_a$ of the $S^j$ alternative.
The local projector will be ${{\cal
P}^j}_a=\int_{R_a}|x,T,+><x,T,+|$, where $|x,T,+>$ are the
eigenstate of ${\hat X(T)},{\hat \epsilon}$ with eigenvalues
$x,+$. $T$ is the proper time when the measurement occurs. These
projectors are defined on each local Lorentz system.  In
principle, we could have different Lorentz time variables on
different regions belonging to the same $S^i$ alternative.  This
procedure allows us to represent the local position measurements
on $S^i$ in terms of local projectors on a covariant Hilbert
space. The projectors associated with different devices on each
$S^j$ commute up to lambda Compton corrections. The reduction
postulate transforms the physical state into the normalized
projection that we have already defined on each set of
alternatives $S^i$.

In general, our description may be extended to any relativistic
quantum theory, like a Dirac particle or even for quantum field
theory. Hence, our approach should be taken as a general framework
for the relativistic domain.  An important consequence is the
following: non-local observables are also well defined in the
relativistic case. In fact, since non-local properties are
measured by means of local observations on a system of measurement
devices separated by space-like intervals \cite{Aha-A}, they may
be included in a set of alternatives $S^i$ and therefore
correspond to a single covariant reduction process. Thus we have
shown that relativistic quantum mechanics admits a natural
realistic interpretation of the quantum states. {\em{Quantum
states are multi-local relational objects that give us the
disposition of the system for producing certain events with
certain probability among a particular and intrinsic set of
alternatives $S^i$}}. The evolution of this disposition is a
well-defined covariant process on the physical Hilbert space in
the generalized Heisenberg picture. The main difference with the
non-relativistic case is that here, in each measurement, the
system provides a result in devices that may be located on
arbitrary space-like surfaces.  It is important to notice that the
above proposed description neither refers to any particular choice
of the evolution parameter nor corresponds to any foliation of
space-time. Therefore, one does not have a natural Schroedinger
picture on this approach.  As it was shown in quantum field theory
\cite{Torre}, the global Schroedinger picture could not exist due
the foliation-dependence of the global state evolution.  However,
as was pointed out by Dirac: ``Heisenberg mechanics is the good
mechanics", and this is also the case here. The only remaining
problem is that the evolving constants defined on a global,
generic curved, Cauchy surface could not be self-adjoint
operators. But since we are working with the standard proper
Lorentzian time of each local measurement device belonging to
$S^i$, the corresponding local operators are self-adjoint. Hence
the relational local point of view adopted here allows us to avoid
this problem.  Nevertheless, the Schroedinger picture is well
defined in a first quantized relativistic theory.  In this
picture, dispositions are associated to multi-local states. Each
multi-local state $\psi_i$ is given by a class whose elements are
the wavefunctions computed on each space-like surface containing
the measurement devices belonging to $S^i$.  One obtains each of
the elements of the class $\psi_i$ by evolving with the wave
equation the initial state.  Reduction takes place on any
particular space-like surface containing the covariant
alternatives.  {\em Notice that contrary to what happens with the
standard Lorentz dependent description, here the conditional
probabilities of further measurements are unique.} It is in that
sense that the dispositions of state to produce further results
has an objective character.
\\

We have developed the relational interpretation adopting the real
tendency theory. Nevertheless, the relational point of view could
be within the context of any realistic interpretation, providing a
covariant reduction process. The main result we have found is that
it is possible to introduce a set of local projectors, covariantly
ordered, and a covariant Hilbert space with in which these
operators are well defined self-adjoint operators and they commute
for a given $S^i$ set. The reduction postulate is now covariantly
defined, this is an important step toward a complete realist
theory. The relational nature of reality should be taken as a
general feature of a relativistic world. In fact, a paradigmatic
example of relational theory is general relativity that
establishes the relational character of space, time and matter.
Space and time are nothing but the dependence of the phenomena on
one other. At the quantum level, systems do not have properties
before the measurement: events are a product of the interaction of
the system with the measurement devices. An even more striking
piece of evidence, in quantum field theory in curved space-time,
the very notion of particle depends upon the motion of the
detectors \cite{Wald}. In that sense, {\em {a system is given by
the set of its behaviors with respect to others}} . An isolated
system does not have properties or attributes, since all the
``properties" result from its interaction with other systems. It
is important to remark that this is a strongly objective
description in the sense of D'Espagnat and it does not make any
reference to operations carried out by human observers
\cite{Esp,DEsp}. Once a quantum system and a set of measurement
devices are given, the evolution of the state is uniquely and
covariantly defined. A final observation is in order. As one can
immediately note, the initial condition has a deep relevance in
the construction of the covariant alternatives. In many cases the
preparation of the system is central for the determination of the
initial condition.  As one already knows, a quantum system involve
entangled objects, therefore in a complete quantum theory one has
to take the whole universe as the system. There, the relational
point of view is the only way for describing the evolution. In
this domain, a quantum object may not have a natural beginning
beyond the big bang. If one is describing a particular portion of
the universe within a given time interval, then one can consider a
partial initial condition given by a particular set of events that
contain the forthcoming alternatives in the forward light cone.
Hence, one naturally falls into a sort of statistical mixture, as
it is the case in non-complete measurements.

\section{Acknowledgments}
The authors thank Michael Reisenberger for helpful suggestions.
This paper was supported in part by PEDECIBA-Uruguay.

\end{multicols}

\end{document}